\documentclass[preprint,5p]{elsarticle}
\usepackage{stmaryrd}
\usepackage{amsmath}
\usepackage{amssymb}
\usepackage{tabularx}
\usepackage{subfigure}
\usepackage{color}
\usepackage{indentfirst}
\SetSymbolFont{stmry}{bold}{U}{stmry}{m}{n}
\biboptions{sort&compress}
\usepackage{graphicx}
\usepackage{enumitem}
\journal{Acta Mater.}
\numberwithin{equation}{section}

\begin{document}

\begin{frontmatter}

\title{Triple junction drag effects during topological changes in the evolution of polycrystalline microstructures}
\author[1]{Quan Zhao\corref{5}}
\address[1]{Department of Mathematics, National University of
Singapore, Singapore, 119076}

\author[2]{Wei Jiang\corref{5}}
\address[2]{School of Mathematics and Statistics {\rm \&} Computational Science
Hubei Key Laboratory, Wuhan University, Wuhan, 430072, China}
\cortext[5]{These two authors contributed equally to this work.}

\author[3]{David J. Srolovitz\corref{4}}
\address[3]{Departments of Materials Science and Engineering {\rm \&}
Mechanical Engineering and Applied Mechanics, University of Pennsylvania, Philadelphia, PA 19104, USA}
\cortext[4]{Corresponding author.}
\ead{srol@seas.upenn.edu}

\author[1]{Weizhu Bao}


\begin{abstract}

Experiments, theory and atomistic simulations show that finite triple junction mobility results in non-equilibrium triple junction angles in evolving polycrystalline systems. These angles have been predicted and verified for cases where grain boundary migration is steady-state. Yet, steady-state never occurs during the evolution of polycrystalline microstructures as a result of changing grain size and topological events (e.g., grain face/edge switching - ``$T_1$'' process, or grain disappearance ``$T_2$'' or ``$T_3$'' processes). We examine the non-steady evolution of the triple junction angle in the vicinity of topological events and show that large deviations from equilibrium and/or steady-state angles occur. We analyze the characteristic relaxation time of triple junction angles $\tau$ by consideration of a pair of topological events, beginning from steady-state migration. Using numerical results and theoretical analysis we predict how the triple junction angle varies with time and how $\tau$ varies with triple junction mobility.  We argue that it is precisely those cases where grain boundaries are moving quickly (e.g., topological process in nanocrystalline materials), that the classical steady-state prediction of the triple junction angle about finite triple junction mobility is inapplicable and may only be applied qualitatively.
\end{abstract}


\end{frontmatter}

\section{Introduction}

In polycrystalline materials, two grains meet at a surface (grain boundary), three grains along a curve (triple line or junction) and four grains at a point (quad point or vertex). In curvature or capillarity-driven grain boundary migration, a grain boundary moves with a velocity that is proportional to its mean curvature~\cite{CS Smith} and the proportionality constant is the reduced mobility (the product of the grain boundary mobility and the grain boundary stiffness)~\cite{Gottstein1998:steadystatemotion}. In the isotropic case, the stiffness is simply the grain boundary energy (per unit area). The curvature is discontinuous along  triple junctions; this implies that the  equilibrium angle (given by the Young-Dupree equation~\cite{young}) is established with infinite velocity.  Hence, in classical grain growth theory, the triple junction angle is usually assumed to be fixed at its equilibrium value and is therefore treated as a boundary condition.

Experimental results (e.g., see~\cite{Czubayko1998:Exp,Gottsteom1999:Exp,
Protasova2001:Exp,Mattissen2005:Exp} )and molecular dynamics simulations (e.g., see~\cite{Upmanyu1999:molecular simulation}), however, show that the triple junction angle needs not be in equilibrium and deviations from equilibrium increase with increasing triple junction velocity.  Similar observations have been made in the case for solid-liquid-vapor, solid-solid-vapor and solid-liquid 1-liquid 2 contact lines during the spreading or retraction of films on substrates~\cite{Moving contact1,Moving contact2,SSD1,SSD2}.  While these observations contradict the constant angle triple junction boundary condition argument above, consideration of the microscopic structure of triple junctions suggest the origin of such deviations from equilibrium. Just like grain boundary mobilities depends on the atomic structure of a grain boundary, so too the triple junction mobility should be expected to depend on its own atomic structure.  While the statement that the triple junction angle is always at its equilibrium value implicitly implies that the triple junction mobility is infinite~\cite{Gottsten2002: theory}, finite triple junction mobilities imply that triple junctions angles will, in general, differ from their equilibrium value for all finite triple junction velocities.

Finite triple junction mobility also implies that triple junctions may provide a drag on grain boundary migration and hence affect microstructure  and grain boundary morphology evolution.  This observation was anticipated in an early theoretical analysis of grain boundary triple junction drag~\cite{Galina1987}. In fact, a series of theoretical ~(e.g., see \cite{Gottstein1998:steadystatemotion,Gottstein2000:theory,Gottsten2002: theory,Gottstein2005:theory}, simulation (e.g., see \cite{Upmanyu1999:molecular simulation}), and experimental (e.g., see~\cite{Czubayko1998:Exp,Gottsteom1999:Exp,Upmanyu1999:molecular simulation,Protasova2001:Exp,Mattissen2005:Exp} studies have focussed on precisely this issue.  In order to simplify the theoretical analysis and the interpretation of experiments, many of these studies were carefully designed to achieve steady-state grain boundary migration profiles ~\cite{Upmanyu2002: study steady} and thus steady-state conditions have been assumed in most analyzes ~\cite{Gottstein1998:steadystatemotion, Gottstein2000:theory,Gottsten2002: theory}.

On the other hand, in most important classes of microstructure evolution (e.g., normal grain growth), grain boundary morphologies do not evolve in a steady-state manner.  For example, the fact that the mean grain size increases during normal grain growth implies that some grains must shrink and disappear. As such grains shrink the mean curvature grows and hence grain boundary velocity is accelerating as the grain shrinks. Formally, this implies that as the grain size of disappearing grain goes to zero, the grain boundary velocity diverges.  This unphysical result in this common limiting case implies both that the classical assumptions of grain boundary migration break down and that  finite triple junction mobility will dominate the evolution as a grain disappears.  In fact, this is true in most cases when topological changes are occurring or, in other words, when triple junctions meet. Such topological events as grain neighbor switching ($T_1$ process) and the disappearance of three- or two-sided grains ($T_2$ or $T_3$ processes) are a central feature of microstructural evolution of grain growth.

In this paper, we focus on the effects of finite triple junction mobility during topological transitions. We do this both because triple junction mobility effects will be pronounced during these common topological processes during grain growth and because it provides a laboratory for observing the effects of finite triple junction mobility in situations where the common assumption of steady-state must fail. In particular, we examine the case where equilibrium triple junction angles change abruptly; namely, during the $T_1$ and $T_3$ topological processes described above. The main goal of this study is to determine the effects of  triple junction drag in non-steady-state conditions and the time scale required for triple junction angles to relax toward those predicted on the basis of steady-state analyses.


\section{Grain boundary dynamics}

In capillarity or grain boundary surface tension-driven grain boundary migration, grain boundary motion is overdamped such that the grain boundary velocity may be written as the product of the grain boundary mobility $m_{\rm{b}}$ and the force on the grain boundary (the variation of the energy with respect to the displacement of the grain boundary) $\sigma\,\kappa$, where   $\sigma$ is the grain boundary surface tension and $\kappa$ is the local mean curvature of the grain boundary. (Note, in this manuscript, we focus explicitly on the classical case where all grain boundaries have equal and isotropic surface tensions and mobilities.)
Following the same approach, we can write the velocity of a triple junction as the product of a mobility $m_{\rm{tj}}$ and the driving force for triple junction motion that arises from the surface tensions of the three grain boundaries meeting there  $\vec{f}_{\rm{tj}}=\sum_{i} \,\sigma\,\vec\tau_i=\sigma\,\vec g_{\rm{tj}}$, where $\vec\tau_i$ is the unit tangent vector along grain boundary $i$ where it joins the triple junction and $\vec g_{\rm{tj}}=\sum_{i}\vec\tau_i$ is its vector sum.
The interaction between the grain boundary and triple junction motions can have a profound influence on the evolution of the entire systems of grain boundaries that comprise the microstructure.
Note here that because the driving force on the grain boundary has the dimensions of a pressure or stress while the driving force on the triple junctions is a force per unit length, the dimensions of the grain boundary and triple junction mobility are different, and the ratio $m_{\rm{b}}/m_{\rm{tj}}$ has the dimension of a length.  Also, note that when the three isotropic grain boundaries meet at the triple junction at an angle of $2\pi/3$ there is no force on the triple junction.
We may write the equations of motion for both the grain boundaries and triple junctions more formally as
\begin{eqnarray}\label{eqn:Bmotion}
\partial_{t} \vec X=m_{b}\,\sigma\,\kappa\,\vec n ,\qquad \kappa=\partial_{ss}\vec X\cdot\vec n,
\end{eqnarray}
and
\begin{equation}\label{eqn:Tmotion1}
\frac{d\vec X_{\rm{tj}}}{dt}=m_{\rm{tj}}\sum_{i} \,\sigma\,\vec\tau_i= m_{\rm{tj}}\,\sigma\,\sum_{i}\vec\tau_i=m_{\rm{tj}}\,\sigma\, \vec{g}_{\rm{tj}},
\end{equation}
where $\vec X:=\vec X(s,t)=(x(s,t),y(s,t))$ represents an arbitrarily curved grain boundary where $s$ and $t$ represent the arc length and time, $\vec X_{\rm {tj}}:=\vec X_{\rm {tj}}(t)=(x_{\rm {tj}}(t),y_{\rm {tj}}(t))$ represents the position of the triple junction, 
and $\vec{n}$ is the outer unit normal vector to the grain boundary. 

\begin{figure}[!t]
\includegraphics[width=8cm,angle=0]{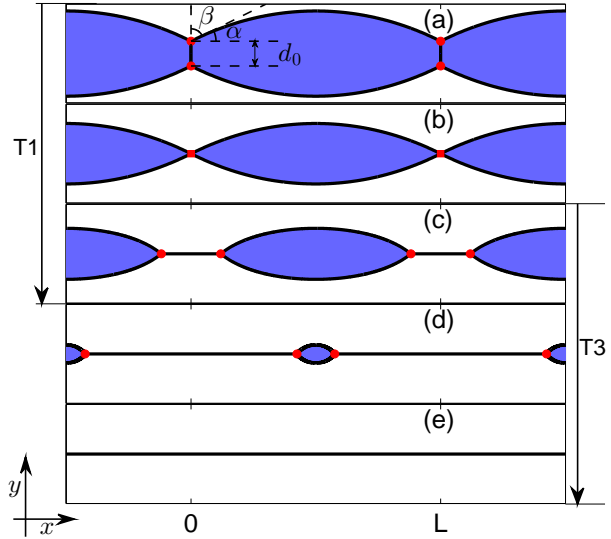}
\caption{A schematic illustration of the temporal evolution of a grain boundary system considered in this paper, which
 includes a $T_1$ process from (a) to (c) and a $T_3$ process from (d) to (e). Here (a) is the steady state motion
during $t<0$; (b) shows the occurring morphology of a $T_1$ process at $t=0$; and (c) $t=t_1^\delta$, (d) $t=t_3^{\delta}$ and (e) $t=t_e^{\delta}$ show several intermediate evolution morphologies during $T_1$ and $T_3$ processes.}
\label{fig:NSSystem}
\end{figure}

For the upper triple junction shown in Fig.~\ref{fig:NSSystem}(a), we can explicitly evaluate the resultant vector in Eq.~\ref{eqn:Tmotion1}, $\vec g_{\rm{tj}}=\sum_{i}\vec\tau_i$: $\tau_1=(0,-1)$, $\tau_2=(\sin\beta,\cos\beta)$ and $\tau_3=(-\sin\beta,\cos\beta)$, resulting in $\vec g_{tj} = (2\cos\beta - 1) \vec e_2$, where $\vec{e}_2$ represents the unit vector along the $y$-direction
and $\beta$ represents the angle as depicted in Fig.~\ref{fig:NSSystem}(a).
As discussed in \cite{Gottstein1998:steadystatemotion},  the above system of equations 
admits a steady-state solution in which the upper GB moves downward (in the $y$-direction) with a velocity of constant magnitude $v_0=m_{\rm tj}\,\sigma(1-2\cos\beta)$ (when $\beta\in(\frac{\pi}{3},\frac{\pi}{2})$),
and the steady-state solution can be written as
\begin{equation}\label{PCurve}
y(x)=\frac{ L}{\pi-2\beta}\left[\ln\cos\frac{(\pi-2\beta)(2x-L)}{2L}
-\ln\sin\beta\right],
\end{equation}
for $0\le x\le L$ with $L>0$ a fixed constant. This solution can be extended
periodically with the period $L$ in $x$-direction for
Eqs.~\eqref{eqn:Bmotion}-\eqref{eqn:Tmotion1}~\cite{Gottstein1998:steadystatemotion}.

We consider the dynamics of the system of grain boundaries shown in Fig.~\ref{fig:NSSystem}(a): initially it consists of two steady-state motions
via the periodic extension of the grain boundary profile in Eq. \eqref{PCurve}, i.e., the upper part of the profile (i.e. $y(x)+d_0/2$ with $d_0>0$ a fixed constant) migrates downwards in  the $y$-direction with  constant velocity $v_0$ and the lower part of the profile (i.e. $-y(x)-d_0/2$) migrates upwards in the $y$-direction with the same constant velocity $v_0$  (see Fig. \ref{fig:NSSystem}(a)). Before the upper and lower parts of the grain boundary profiles meet, the triple junction angle must be between $\frac{\pi}{3}$ and $\frac{\pi}{2}$.

\begin{figure*}[!htp]
\centering
\includegraphics[width=16.0cm,angle=0]{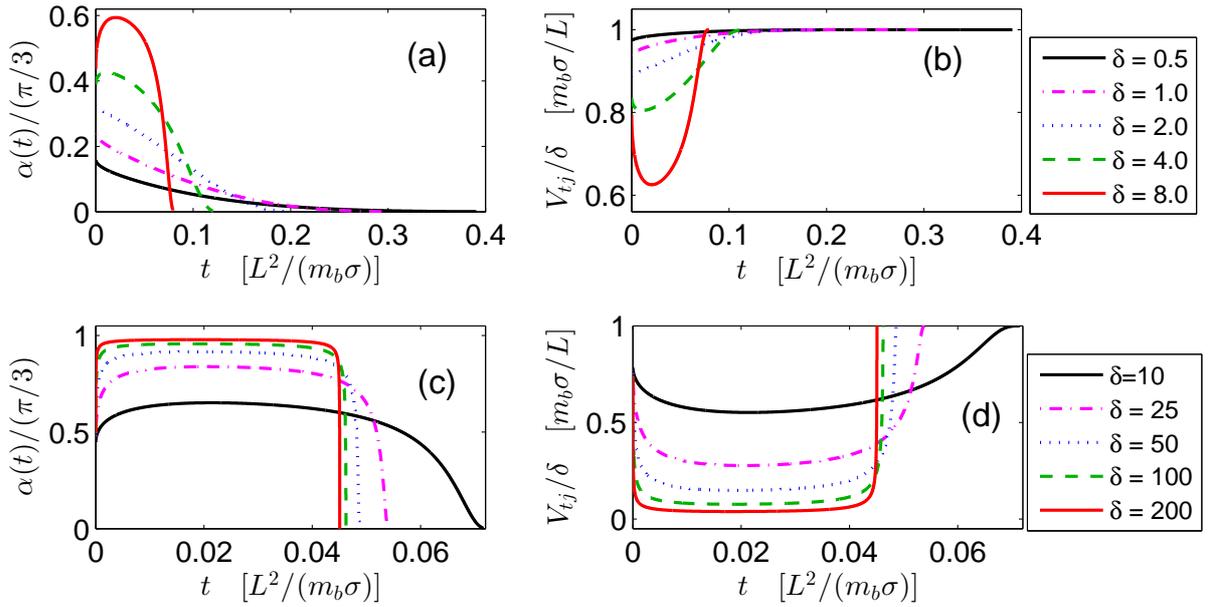}
\caption{Temporal evolution of the triple junction angle $\alpha(t)$ (left column) and the magnitude of triple junction velocity $V_{\rm tj}(t)$ (right column) immediately following a $T_1$ event (at $t=0$) for  small (a and b) and large (c and d) values of the mobility ratio parameter $\delta>0$.}
\label{fig:TPavstj}
\end{figure*}

The upper and lower sections of the grain boundary profile move toward each other until they meet at a time we designate $t=0$, for convenience  (cf.~Fig.~\ref{fig:NSSystem}(b)). Once these two triple junctions meet, a topology change event will occur, i.e., the junction will subsequently split into two different triple junctions and abruptly migrate away from each other in the
$\pm x$-direction (instead of the $\pm y$-direction), as shown in Fig.~\ref{fig:NSSystem}(c)). This topology change event,
i.e., the switching of which grains are neighbors of each other is called a $T_1$ process~\cite{D.Weaire T1 process, D. Weygand T3 Process, Moldovan02}.
Note that during the $T_1$ process, the two triple junctions change their direction of motion by $\pm \frac{\pi}{2}$ (cf.~Fig.~\ref{fig:NSSystem} (a) and (b)).
This abrupt change of the triple junction angle, i.e., from $\beta$ to $\alpha:=\alpha(t)$ with $0< \alpha(t=0)=\frac{\pi}{2}-\beta< \frac{\pi}{6}$ will instantly change the resultant vector on the triple junction from $\vec{g}_{\rm{tj}}=(2\cos\beta-1)\;\vec{e}_2$ to $\vec{g}_{\rm{tj}}=(2\cos\alpha-1)\;\vec{e}_1$ with $\vec{e}_1$  the unit vector along the $x$-direction. The abrupt change in the direction of triple junction motion and the steady-state triple junction angle converts the grain boundary dynamics from a steady-state to a non-steady state motion.

After the $T_1$ process occurs, the grain boundaries will continue to evolve via mean curvature flow, and the grains (shaded in blue in Fig.~\ref{fig:NSSystem}(d) will become smaller and smaller) until they eventually disappear and form a straight line (see Fig.~\ref{fig:NSSystem}(e)). The final disappearance of the shaded grain is a $T_3$ process~\cite{Moldovan02}.

To aid the discussion (and numerical analysis) below, we perform the following changes of variables: the lengths and time variables
are, respectively, normalized by the length parameter $x_s=L$ and the time parameter $t_s={L^2}/{(m_b\sigma)}$,
where $L$ represents the periodic length of the grain boundary steady-state profile defined in Eq.~\eqref{PCurve} (also shown in Fig.~\ref{fig:NSSystem}). With the above set of non-dimensional variables (we still use the same notations for brevity), Eqs.~\eqref{eqn:Bmotion}-\eqref{eqn:Tmotion1} can be non-dimensionalized as follows:
\begin{equation}\label{boundary_motion}
\partial_{t}\vec X=\kappa\;\vec n,\qquad \kappa=\partial_{ss}\vec X\cdot\vec n,
\end{equation}
\begin{equation}\label{triple_motion}
\frac{d\vec X_{\rm{tj}}}{dt}=\delta\,\vec{g}_{\rm{tj}},
\end{equation}
where $\delta$ is the dimensionless parameter
\begin{equation}
\delta=\frac{L\,m_{\rm{tj}}}{m_b}>0.
\end{equation}

In fact, the steady-state triple junction angle $\beta$ and the dimensionless parameter $\delta$ are related as follows~\cite{Gottstein1998:steadystatemotion}:
\begin{equation}\label{PTheta}
2\beta-2\delta\cos\beta+\delta -\pi=0.
\end{equation}
This suggests that $\beta(\delta)$ is a monotonically decreasing function
for $\delta\ge0$. Asymptotic analysis shows that
\begin{equation}\label{eqn:alphaasym}
\beta\approx \left\{\begin{array}{ll} \frac{\pi}{2}-\frac{\delta}{2}+\mathcal{O}(\delta^2),
&\quad\text{for}\;0<\delta\ll 1; \\
\frac{\pi}{3}+\frac{\pi}{3\sqrt{3}\delta}+\mathcal{O}(\delta^{-2}),
&\quad\text{for}\;\delta\gg 1.\\
\end{array}\right.
\end{equation}
In addition, from these relations, we see that $\beta\to \frac{\pi}{2}$ as  $\delta\to 0$ and $\beta\to\frac{\pi}{3}$ as  $\delta\to\infty$.

In the following, we  focus on how triple junction drag affects grain boundary motion during  $T_1$ and $T_3$ processes. For simplicity, we assume that the grain boundary migration shown in Fig.~\ref{fig:NSSystem}(a)) for $t<0$ is in steady-state and all of the results presented below start at the time ($t=0$) (see Fig.~\ref{fig:NSSystem}(b)-(e)).

\section{Drag effect during $T_1$ and $T_3$ processes}

We used a parametric finite element method~\cite{Barret2008:PFEM, Bao16} to solve the above sharp-interface model, i.e., Eqs.~\eqref{boundary_motion}-\eqref{triple_motion}. We focus on grain boundary dynamics from the steady-state
profile (see Fig.~\ref{fig:NSSystem}(a) and Eq.~\eqref{PCurve}) through the subsequent non-steady state motion ($t\geq0$)  - all length units are scaled by $L$ and hence the results are independent of the discretization.  Figure~\ref{fig:TPavstj} shows the temporal evolution of the triple junction angle $\alpha(t)$ 
and the triple junction velocity $V_{tj}(t)$ 
for $0.5\leq\delta\leq200$.
These numerical results demonstrate that the dimensionless parameter $\delta$ has profound effects on the evolution of the triple junction angle $\alpha(t)$.

\begin{figure}[htp]
\centering
\includegraphics[width=8.0cm,angle=0]{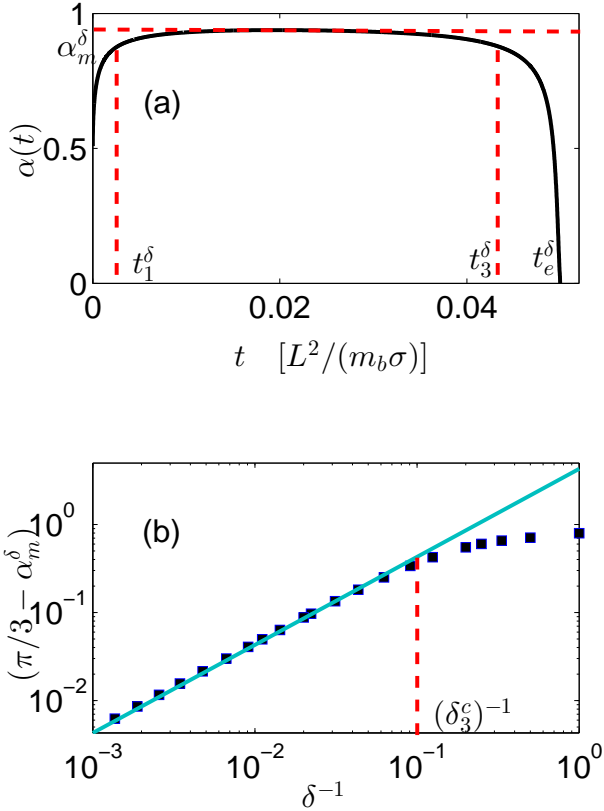}
\caption{(a) Plot of $\alpha(t)$ for $\delta=20$ and
(b) numerical results for $\alpha_m^\delta:=\max_{t\ge0} \alpha(t)$ for
different $\delta\gg1$ (symbols) and its asymptotic fitting
behavior (cyan solid line)
$\alpha_m^\delta=\frac{\pi}{3}-4.3 \delta^{-1}$ when $\delta\ge\delta_3^c\approx 10$.}
\label{fig:Max_alpha}
\end{figure}

By our extensive numerical results (see Fig.~\ref{fig:TPavstj}), according to the magnitude of the dimensionless parameter $\delta$, we can categorize the dynamic evolution process of the triple junction angle $\alpha(t)$ into the following three different cases: (i) when $0< \delta < \delta_1^c \approx 2.5$, the triple junction angle $\alpha(t)$ will decrease slowly and monotonically  from its initial value $\alpha(0)$ to zero (see Fig.~\ref{fig:TPavstj}(a)); (ii) when $\delta_1^c < \delta <\delta_2^c\approx10$, the triple junction angle $\alpha(t)$ will first increase to a maximum value and then decrease quickly to zero (see Fig.~\ref{fig:TPavstj}(a)); (iii) when $\delta>\delta_2^c$, the triple junction angle $\alpha(t)$ will first increase rapidly from its initial value $\alpha(0)$ to a maximum value $\alpha_m^{\delta}$ at time $t=t_1^\delta$, then maintain this maximum value as a plateau until $t=t_3^\delta$, and finally decrease rapidly to zero at $t=t_e^\delta$ (see Fig. \ref{fig:Max_alpha}(a)).

The rate of change of the grain area $S:=S(t)$ for the above system in one period for $t\geq 0$ can be calculated
analytically as:
\begin{equation}{\label{eqn:Area_shrink}}
\frac{dS}{dt}=-\oint v_n ds=-4\alpha(t), \quad 0\leq \alpha(t) \leq \frac{\pi}{3},
\end{equation}
where $v_n$ is the magnitude of normal velocity of the grain boundary. From the above equation, we can see that
the triple junction angle $\alpha(t)$ is a good indicator which can be used to represent the drag effects which are exerted by triple junctions on grain boundary migration. In addition, when $\delta$ is large, $\alpha(t)$ approaches $\pi/3$ (i.e., it approaches its equilibrium value and the relaxation time for $\alpha$ is small, see Fig.~\ref{fig:TPavstj}). In this limit, $\frac{dS}{dt}\approx-\frac{4\pi}{3}$ and the classical  von Neumann-Mullins relation will be valid, i.e., $\frac{dS}{dt}=\frac{\pi}{3}(n-6)$ for a grain growth where $n$ is the number of triple junctions (where $n=2$).

\begin{table}[htp]
\caption{Numerical fitting results for $p(\delta)$, $q(\delta)$,
$\tau_1(\delta)$ and $t_1^\delta$ in the fitting formula Eq.~\eqref{eqn:T1equation} during the $T_1$ process for different $\delta\gg1$. }
\medskip
\def\temptablewidth{0.5\textwidth}
\vspace{-12pt}
{\rule{\temptablewidth}{1pt}}
\begin{tabular*}{\temptablewidth}{@{\extracolsep{\fill}}ccccc}
$\delta$  &$p(\delta)$ &$q(\delta)$   &$\tau_1(\delta) [\frac{L^2}{m_b\sigma}] $
&$t_1^{\delta}  [\frac{L^2}{m_b\sigma}]$\\ \hline
20  &$0.197$  &$0.211$ &$1.27\times 10^{-3}$ &$5.95\times10^{-3}$ \\\hline
30  &$0.143$   &$0.178$   &$7.26\times 10^{-4}$     &$4.45\times10^{-3}$\\ \hline
50	&$0.102$   &$0.152$	&$3.18\times 10^{-4}$	 &$2.75\times 10^{-3}$\\ \hline
70   &$0.0877$ &$0.127$ &$1.57\times10^{-4}$ &$1.89\times 10^{-3}$\\\hline
100	&$0.0712$   &$0.130$	&$9.35\times 10^{-5}$	 &$1.20\times 10^{-3}$\\ \hline
150	&$0.0609$	&$0.117$	&$4.41\times 10^{-5}$	 &$6.74\times 10^{-4}$\\ \hline
200	&$0.0555$	&$0.106$	&$2.55\times 10^{-5}$	 &$4.33\times 10^{-4}$\\ \hline
400	&$0.0462$	&$0.0696$	&$6.76\times 10^{-6}$	 &$1.37\times 10^{-4}$\\ \hline
800	&$0.0352$	&$0.0234$	&$1.98\times 10^{-6}$	 &$3.94\times 10^{-5}$\\
\end{tabular*}
{\rule{\temptablewidth}{1pt}}
\label{tab:t1}
\end{table}

For Case (iii) (i.e., $\delta>\delta_2^c\approx 10$), the triple junction angle $\alpha(t)$ rapidly increases from its initial value $\alpha(0)$ to a maximum value $\alpha_m^\delta$ during the $T_1$ process (see Fig.~\ref{fig:Max_alpha}(a)), i.e., when two grains lose a side and two others gain a side.
This neighbor switching event implies that the triple junction angle will immediately change from $\beta\approx \frac{\pi}{3}$ to $\alpha(0)=\frac{\pi}{2}-\beta\approx \frac{\pi}{6}$ (for sufficiently large $\delta$) and the two newly separated triple junctions will migrate in the $\pm x$ directions, respectively.
Our numerical  results show that  large deviation of the triple junction angle from $\pi/3$ to about $\pi/6$ leads to a rapid increase in the triple junction angle $\alpha(t)$ to $\alpha^\delta_m\approx \frac{\pi}{3}$ from its initial value $\alpha(t=0)$ over  $0\leq t \leq t_1^\delta$. We find that the numerical results for the time evolution of the triple junction angle $\alpha(t)$ during this process is well fitted by the following relation
\begin{equation}
\alpha(t)=\frac{\pi}{3}-p(\delta)-\frac{\frac{\pi}{6}-q(\delta)}{1+ t/\tau_1(\delta)}, \quad  0\leq t \leq t_1^{\delta},
\label{eqn:T1equation}
\end{equation}
where $p(\delta)$, $q(\delta)$ and $\tau_1(\delta)$ are tabulated in Table~1 for large $\delta$ (i.e., Case (iii)).

\begin{table}[htp]
\caption{Numerical fitting results for $r(\delta)$,
$\tau_3(\delta)$ and numerical results for $t_e^\delta-t_3^\delta$, $t_e^{\delta}$ for different
$\delta\gg1$ in the fitting formula Eq.~\eqref{eqn:T3equation}
during the  $T_3$ process.}
\def\temptablewidth{0.5\textwidth}
\medskip
\vspace{-12pt}
{\rule{\temptablewidth}{1pt}}
\begin{tabular*}{\temptablewidth}{@{\extracolsep{\fill}}ccccc}
$\delta$  &$r(\delta)$    &$\tau_3(\delta)[\frac{L^2}{m_b\sigma}] $   &$(t_e^{\delta}-t_3^{\delta})[\frac{L^2}{m_b\sigma}]$  &$t_e^{\delta} [\frac{L^2}{m_b\sigma}]$\\ \hline
20 &$0.0856$  &$3.07\times10^{-3}$ &$1.54\times10^{-2}$ &$0.0566$\\\hline
30 &$0.0838$    &$1.17\times10^{-3}$     &$1.05\times10^{-2}$  &$0.0521$\\\hline
50	& $0.0775$	&$4.07\times10^{-4}$	&$6.38\times10^{-3}$   &$0.0487$\\\hline
70 &$0.0729$ &$2.09\times10^{-4}$ &$4.49\times10^{-3}$ &$0.0473$\\\hline
100	&$0.0688$	&$1.01\times10^{-4}$	&$2.94\times10^{-3}$  &$0.0463$\\\hline
150	&$0.0648$	&$4.48\times10^{-5}$	&$1.74\times10^{-3}$  &$0.0455$\\ \hline
200	&$0.0626$	&$2.51\times10^{-5}$	&$1.16\times10^{-3}$  &$0.0451$\\ \hline
400	&$0.0598$	&$6.10\times10^{-6}$	 &$3.88\times10^{-4}$  &$0.0445$\\ \hline
800	&$0.0584$	&$1.46\times10^{-6}$	&$1.15\times10^{-4}$    &$0.0444$\\
 \end{tabular*}
{\rule{\temptablewidth}{1pt}}
\label{tab:t3}
\end{table}

\begin{figure}[htp]
\centering
\includegraphics[width=8.0cm]{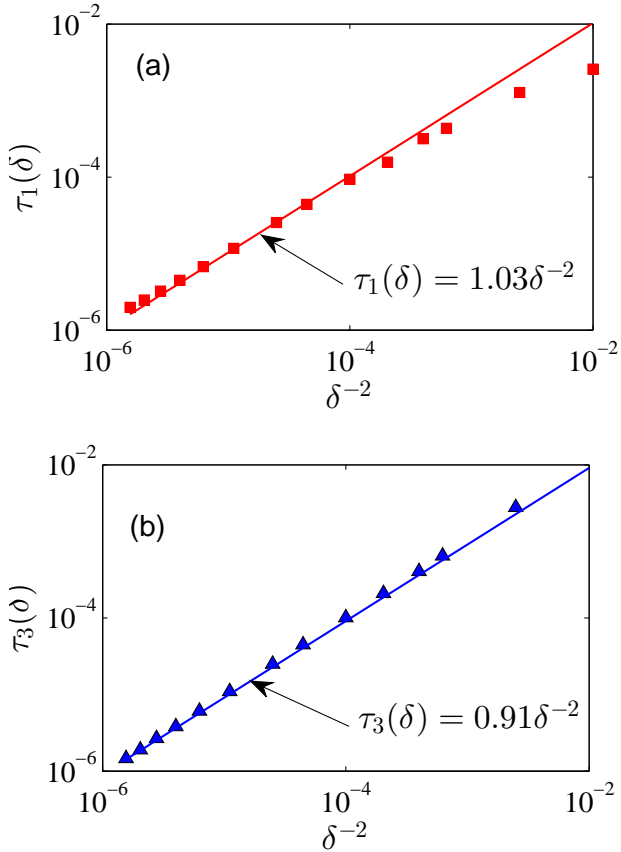}
\caption{(a). Plot of $\tau_1(\delta)$ (symbols for numerical results) and
its asymptotic fitting formula $\tau_1(\delta)=1.03\delta^{-2}$
(solid line) for different $\delta$ during the $T_1$ process
and (b). plot of $\tau_3(\delta)$ (symbols for numerical results) and
its asymptotic fitting formula $\tau_3(\delta)=0.91\delta^{-2}$
(solid line) for different $\delta$ during the $T_3$ process.}
\label{fig:mainresult}
\end{figure}

As Table~1 shows, $p(\delta)$, $q(\delta)$, $t_1^\delta$ and $\tau_1(\delta)$ are monotonically decreasing functions of $\delta$ . In addition, our numerical results show that (see Fig.~\ref{fig:mainresult}(a))
\[p(\delta)\approx 0, \quad q(\delta)\approx 0,
\quad \tau_1(\delta)\approx 1.03\delta^{-2}, \quad t_1^\delta \approx 18.4\delta^{-2},\]
when $\delta\gg1$. This immediately implies that, when $\delta$ is very large,
Eq.~\eqref{eqn:T1equation} collapses to the following relation:
\begin{equation}
\alpha(t)=\frac{\pi}{3}-\frac{\frac{\pi}{6}}
{1+t/\tau_1(\delta)}, \quad  0\leq t \leq t_1^{\delta},
\label{eqn:T1new}
\end{equation}
where the relaxation time $\tau_1(\delta) \approx 1.03\delta^{-2}$.

For Case (iii) (i.e. $\delta>\delta_2^c$), we also observe that during the $T_1$ process, the triple junction angle reaches the plateau at $\alpha_m^{\delta}$ and the triple junction velocity reaches the plateau at  $V_m^\delta \approx\delta(2\cos\alpha_m^\delta-1)$ for $t_1^\delta <t<t_3^\delta$ (see Fig.~\ref{fig:NSSystem} and \ref{fig:Max_alpha}(a)). Fig.~\ref{fig:Max_alpha}(b) shows how the triple junction angle plateau $\alpha_m^\delta$ depends on $\delta$. By numerical simulations, we obtained the following relations for the above quantities:
\[ 
\alpha_m^\delta \approx \frac{\pi}{3}-4.3\delta^{-1},
\quad V_m^\delta\approx7.45-9.25\delta^{-1}, \quad \delta\ge \delta_3^c\approx 10.
\]

Subsequently, the two newly formed triple junctions during the $T_1$ process will approach each other and eventually annihilate in the $T_3$ process (see Fig.~\ref{fig:NSSystem}(c)-(e)). In Case (iii), we see that the triple junction angle $\alpha(t)$ begins to rapidly decrease immediately before the $T_3$ event (i.e., $t\ge t_3^\delta$),  eventually decreasing to zero as the triple junction drag effect  becomes increasingly pronounced  as the enclosed grain becomes small, the grain boundary curvature increases and the triple junction velocity rapidly grows (see Fig.~\ref{fig:TPavstj}(c)-(d)).  We can capture the non-steady state triple junction angle evolution which is inherent to the $T_3$ process by
\begin{equation}
\alpha(t)=\frac{\pi}{3}-r(\delta)-\frac{\frac{\pi}{3}-r(\delta)}
{1+(t_e^{\delta}-t)/\tau_3(\delta)}, \quad t_3^{\delta}\leq t\leq t_e^{\delta},
\label{eqn:T3equation}
\end{equation}
where $r(\delta)$, $\tau_3(\delta)$, $t_e^\delta$ and $t_3^\delta$ are displayed in Table~\ref{tab:t3}.
As clearly shown in Table~2, $r(\delta)$, $t_3^\delta$ ,$t_e^\delta$ and $\tau_3(\delta)$ are monotonically decreasing functions. In addition, when $\delta\gg1$,
we obtain numerically (see Fig.~\ref{fig:mainresult}(b))
\[r(\delta)\approx 0,\;\tau_3(\delta)\approx 0.91\delta^{-2},\; t_e^\delta-t_3^\delta\approx 49.9\delta^{-2}, \;t_e^\delta\approx 0.0444.\]
This immediately implies that, when $\delta$ is very large, the relation~\eqref{eqn:T3equation} collapses to the following relation:
\begin{equation}
\alpha(t)=\frac{\pi}{3}-\frac{\frac{\pi}{3}}
{1+(t_e^{\delta}-t)/\tau_3(\delta)}, \quad t_3^{\delta}\leq t\leq t_e^{\delta},
\label{eqn:T3new}
\end{equation}
where the relaxation time $\tau_3(\delta) \approx 0.91\delta^{-2}$.

\section{Discussion and Conclusions}

Experiments and atomistic simulations show that the standard assumption that triple junction angles are always in equilibrium fails when grain boundaries move quickly, such as in nanoscale materials where grain sizes are very small.  This effect is now understood to be attributable to finite triple junction mobility (equilibrium triple junction angles imply infinite triple junction mobility). Many experimental observations and theoretical and simulation studies focus on the effects of triple junction mobility under conditions where the triple junction velocity is constant; i.e., steady-state migration.  On the other hand, grain boundary network evolution in real microstructures is fastest immediately proceeding or following topological events (i.e., where grain boundary curvature is especially large).  Hence, the effects of finite triple junction mobility or triple junction are most pronounced near topological transition and such transitions are an essential feature of microstructure evolution in materials with both nano-macro-scale mean grain sizes. In this study, we focussed on the essentially non-steady-state case of triple junction motion/grain boundary migration that accompanies topological transitions.

In particular, we studied triple junction drag effects during the $T_1$ process (neighbor grain switching) and the $T_3$ process (disappearance of two-sided grains). To connect with earlier work, we devised a simulation geometry where the migration is steady-state before and after these topological events (but not too near the events themselves). For the most important case, where the triple junction mobility is large, but finite, we identified a dimensionless parameter $\delta=\frac{L\,m_{\rm{tj}}}{m_b}$  that plays a key  role in describing the dynamical evolution of the triple junction angle and triple junction velocity during the $T_1$ and $T_3$ processes, where $L$ the characteristic length of the problem and $m_{\rm{tj}}$ and $m_b$ are the triple junction and grain boundary mobilities (as in many of the steady-state analyses).

The key results of our study is the prediction of the conditions under which consideration of triple junction mobility near topological events is essential and the  time-scale $\tau$ over which such effects persist.  The triple junction angle varies with time in the vicinity of a topological event as $\alpha(t)=A-B/(1\pm t/\tau)$ where the constants $A$ and $B$ depend on the steady-state triple junction angle before and after the topological event.  The $\pm$ refers to whether the variation in the angle is after or before the topological event. Our results demonstrate that the relaxation time associated with the non-steady-state triple junction angle scales as $\tau\sim\delta^{-2}$.  It is important to note, that the variation of the triple junction angle discussed here is in addition to the well-studied dependence of triple junction angle on triple junction mobility.

While the variation of triple junction angle in the vicinity of topological events is a general feature of microstructure evolution and finite triple junction mobility, the form of this variation and the relaxation to  steady- (or near-steady) state triple junction angles also depends on the magnitude of $\delta$.  While the main results summarized above are valid for $\delta >\delta_2^c\approx10$, for smaller $\delta$ (i.e., $\delta < \delta_2^c$ - small triple junction mobilities) steady-state triple junction angles are never achieved and the classical steady-state prediction of the finite triple junction mobility triple junction angle is never applicable.  Hence, the classical predictions of triple junction angle and its dependence on triple junction angle is only appropriate above some threshold value of $\delta$ and hence should be used judiciously in cases of microstructure evolution where the grain boundaries are moving quickly (e.g., the often discussed case of nanocrystalline materials).

\section*{Acknowledgements}
This work was supported by the National Natural Science Foundation of China Nos. 11401446, 11671312 and 91630313 (W.J.), the Academic Research Fund of the Ministry of Education of Singapore grant No. R-146-000-223-112 (Q.Z.\& W.B.) and as part of the Center for the Computational Design of Functional Layered Materials, an Energy Frontier Research Center funded by the U.S. Department of Energy (DOE), Office of Science, Basic Energy Sciences (BES), under Award \# DE-SC0012575 (D.J.S.). W.J., D.J.S. and W.B. designed research, developed models and wrote the paper, and Q.Z. performed numerical simulations and analyzed data.



\section*{References}
\begin{thebibliography}{s99}
{
\bibitem{CS Smith}
 C.S.~Smith, Introduction to grains, phases, and interfaces: an interpretation of microstructure, Trans. AIME. 175 (1948) 15-51.

\bibitem{Gottstein1998:steadystatemotion}
G.~Gottstein, L.S.~Shvindlerman, Triple junction dragging and von Neumann-Mullins relation, Scr. Mater. 38 (1998) 1541-1547.

\bibitem{young}
T.~Young, An essay on the cohesion of fluids, Philos. Trans. R. Soc. Lond. 95 (1805) 65-87.

\bibitem{Czubayko1998:Exp}
U.~Czubayko, V.G.~Sursaeva, G.~Gottstein, L.S.~Shvindlerman, Influence of triple junctions on grain boundary motion, Acta Mater. 46 (1998) 5863-5871.

\bibitem{Gottsteom1999:Exp}
G.~Gottstein, V.~Sursaeva, L.S.~Shvindlerman, The effect of triple junctions on grain boundary motion and grain microstructure evolution, Interf. Sci. 7 (1999) 273-283.

\bibitem{Protasova2001:Exp}
S.G.~Protasova, G.~Gottstein, D.A.~Molodov, V.G.~Sursaeva, L.S.~Shvindlerman, Triple junction motion in aluminum tricrystals,
Acta Mater. 49 (2001) 2519-2525.

\bibitem{Mattissen2005:Exp}
D.~Mattissen, D.A.~Molodov, L.S.~Shvindlerman, G.~Gottstein, Drag effect of triple junctions on grain boundary and grain growth kinetics in aluminium,
Acta Mater. 53 (2005) 2049-2057.

\bibitem{Upmanyu1999:molecular simulation}
M.~Upmanyu, D.J.~Srolovitz, L.S.~Shvindlerman, G.~Gottstein, Triple junction mobility: a molecular dynamics study, Interf. Sci. 7 (1999) 307-309.

\bibitem{Galina1987}
A.V.~Galina, V.Y.~Fradkov, L.S.~Shvindlerman, Influence of grain ternary joint mobility
on boundary migration, Phys. Met. Metall. 63 (1987) 1220-1222.

\bibitem{Moving contact1}
P.d.~Gennes, Wetting: statics and dynamics, Rev. Mod. Phys. 57 (1985) 827.

\bibitem{Moving contact2}
W.~Ren, W.~E, Contact line dynamics on heterogeneous surfaces, Phys. Fluids. 23 (2011) 072103.


\bibitem{SSD1}
Y.~Wang, W.~Jiang, W.~Bao, D.J.~Srolovitz, Sharp interface model for solid-state dewetting problems with weakly anisotropic surface energies, Phys. Rev. B 91 (2015) 045303.


\bibitem{SSD2}
W.~Jiang, Y.~Wang, Q.~Zhao, D.J.~Srolovitz, W.~Bao, Solid-state dewetting and island morphologies in strongly anisotropic materials , Scr. Mater. 115 (2016) 123-127.


\bibitem{Gottsten2002: theory}
G.~Gottstein, L.S.~Shvindlerman, Triple junction drag and grain growth in 2D polycrystals, Acta Mater. 50 (2002) 703-713.


\bibitem{Gottstein2000:theory}
G.~Gottstein, A.H.~King, L.S.~Shvindlerman, The effect of triple-junction drag on grain growth, Acta Mater. 48 (2000) 397-403.

\bibitem{Gottstein2005:theory}
G.~Gottstein, Y.~Ma, L.S.~Shvindlerman, Triple junction motion and grain microstructure evolution, Acta Mater. 53 (2005) 1535-1544.


\bibitem{Upmanyu2002: study steady}
M.~Upmanyu, D.J.~Srolovitz, L.S.~Shvindlerman, G.~Gottstein, Molecular dynamics simulation of triple junction migration,
Acta Mater. 50 (2002) 1405-1420.


\bibitem{Grainsize}
L.A.~Barrales-Mora, G.~Gottstein, L.S.~Shvindlerman, Effect of a finite boundary junction mobility on the growth rate of grains in two-dimensional polycrystals, Acta Mater. 60 (2012) 546-555.

\bibitem{D.Weaire T1 process}
D.~Weaire, J.~Kermode, Computer simulation of a two-dimensional soap froth. I. Method and motivation, Phil. Mag. B 48 (1983) 245-259.

\bibitem{D. Weygand T3 Process}
D.~Weygand, Y.~Brechet,  J.~Lepinoux, A vertex dynamics simulation of grain growth in two dimensions, Phil. Mag. B 78 (1998) 329-352.

\bibitem{Moldovan02}
D.~Moldovan, D.~Wolf, S.R.~Phillpot, A.J.~Haslam, Mesoscopic simulation of two-dimensional grain growth with anisotropic grain-boundary properties, Phil. Mag. A 82 (2002) 1271-1297.

\bibitem{Barret2008:PFEM}
J.W.~Barrett, H.~Garcke, N.~Robert,  Parametric approximation of Willmore flow and related geometric evolution equations, SIAM J. Sci. Comput. 31 (2008) 225-253.

\bibitem{Bao16}
W.~Bao, W.~Jiang, Y.~Wang, Q.~Zhao, A parametric finite element method for solid-state dewetting problems with anisotropic surface energies, J. Comput. Phys. 330 (2017) 380-400.
}
\end{thebibliography}
\end{document}